\documentclass[aps,preprint,amsmath,amssymb]{revtex4}

\usepackage{graphicx}
\begin{document}

\title{Lepton flavor violation in
medium energy setup of a beta-beam facility}

\author{\.{I}. \c{S}ahin}
\email[]{inancsahin@karaelmas.edu.tr}
 \affiliation{Department of Physics, Zonguldak Karaelmas University, 67100 Zonguldak, Turkey}

\author{M. K\"{o}ksal}
\email[]{mkoksal@cumhuriyet.edu.tr} \affiliation{Department of
Physics, Zonguldak Karaelmas University, 67100 Zonguldak, Turkey}
\affiliation{Department of Physics, Cumhuriyet University, 58140
Sivas, Turkey}

\begin{abstract}
We examine lepton flavor violating couplings of the neutrinos to $W$
boson in medium energy setup of a beta-beam experiment. We show that
muon production via quasielastic scattering and deep inelastic
scattering processes $\nu_e n\to \mu^-\, p$ and $\nu_e N\to \mu^- X$
are very sensitive to $W\mu\nu_e$ couplings. We perform a model
independent analysis and obtain $95\%$ confidence level bounds on
these couplings.
\end{abstract}

\maketitle

\section{Introduction}

Lepton number $(L_i=L_e,L_\mu,L_\tau)$ is conserved in the standard
model (SM) with massless neutrinos. However experimental data such
as those coming from Super-Kamiokande \cite{Kamiokande1,Kamiokande2}
or Sudbury Neutrino Observatory \cite{Ahmad:2002jz}  showed evidence
of neutrino oscillations. These experimental results imply lepton
flavor violation (LFV) and the lepton sector of the SM requires an
extension including massive neutrinos. Since conservation of the
lepton flavor number is not a law of nature that we rely on, it is
significant to quest for new physics via lepton flavor violating
effective operators. Effective lagrangian extension of the SM  could
have an impact on particle physics and astrophysics. Therefore it is
crucial to investigate the physics potential of neutrino experiments
to probe LFV.

Neutral-current lepton flavor violating processes such as $Z\to
\ell_i^\pm \ell_j^\mp$ $(\ell_i=e,\mu,\tau)$ can be observable at
colliders and effective operators contributing to these processes
have been stringently constrained by collider experiments
\cite{DiazCruz:1999xe,FloresTlalpa:2001sp}. On the other hand, in
colliders neutrinos are not detected directly in the detectors.
Instead, their presence is inferred from missing energy signal.
Therefore it is impossible to detect a neutrino flavor at a collider
and charged-current processes such as $W^\pm\to \ell_i^\pm \nu_j$
can not be discerned. In order to constrain lepton flavor violating
couplings of neutrinos to $W$ boson, beta beams provides an
excellent opportunity. Beta beams are electron neutrino and
antineutrino beams produced via the beta decay of boosted
radioactive ions \cite{Zucchelli:2002sa}. Such decays produce pure,
intense and collimated neutrino or antineutrino beams. Purity of the
neutrino beam make it possible to probe couplings $W\ell_i \nu_e$
with a high precision.

In the original beta beam scenario ion beams are accelerated in the
proton synchrotron or super proton synchrotron at CERN up to a
Lorentz gamma factor of $\gamma \sim 100$, and then they are allowed
to decay in the straight section of a storage ring. After the
original proposal, different options for beta beams were
investigated. A low gamma ($\gamma =5-14$) option was first proposed
by Volpe \cite{Volpe:2003fi}. Physics potential of low-energy beta
beams was discussed in detail. It was shown that such beams could
have an important impact on nuclear physics, particle physics and
astrophysics
\cite{McLaughlin:2003yg,Serreau:2004kx,McLaughlin:2004va,Volpe:2005iy,
Balantekin:2005md,Balantekin:2006ga,Volpe:2006in,Jachowicz:2006xx,
Bueno:2006yq,Lazauskas:2007va,Amanik:2007zy,Jachowicz:2008kx}.
Higher gamma options for the beta beams have also been studied in
the literature \cite{Volpe:2006in,BurguetCastell:2003vv,
BurguetCastell:2005pa, Migliozzi:2005zh, Donini:2006tt,
Terranova:2004hu, Huber:2005jk, Donini:2005qg,
Donini:2004hu,Mezzetto:2003ub,Balantekin:2008vw}. A higher gamma
factor provides several advantages. Firstly, neutrino fluxes
increase quadratically with the gamma factor. Secondly, neutrino
scattering cross sections grow with the energy and hence
considerable enhancement is expected in the statistics. An
additional advantage of a higher gamma option is that it provides us
the opportunity to study deep inelastic neutrino scattering from the
nucleus. Very high gamma ($\sim 2000$) options would require
modifications in the original plan such as using LHC and therefore
extensive feasibility study is needed. In this context medium energy
setup is more appealing and less speculative.

In this paper we investigate the physics potential of a medium
energy setup ($\gamma =350 - 580$) proposed in Ref.
\cite{BurguetCastell:2003vv} to probe lepton flavor violating
couplings $W\mu\nu_e$. As far as we know, phenomenology of LFV in a
beta beam facility was studied only in Ref. \cite{Agarwalla:2006ht}
but considering supersymmetric models. Different from
\cite{Agarwalla:2006ht} we have probed LFV in a model independent
way by means of the effective Lagrangian approach.

The organization of this paper is as follows: In the next section we
outline the effective Lagrangian approach. In section III we
summarize the neutrino fluxes and the cross sections for
quasielastic and deep inelastic scattering and present our main
results. Finally Section IV includes concluding remarks.

\section{Effective Lagrangian for lepton flavor violating $W\ell_i \nu_j$ couplings}

There is an extensive literature on non-standard interactions of
neutrinos
\cite{Balantekin:2008vw,Davidson:2003ha,Bell:2005kz,Ibarra:2004pe,Mohapatra:2005wg,Mohapatra:2006gs,de
Gouvea:2007xp,Perez:2008ha,Balantekin:2008rc,Balantekin:2008ib,Biggio:2009nt}.
New physics contributions to lepton flavor violating $W\ell_i \nu_j$
couplings can be investigated in a model independent fashion by
means of the effective Lagrangian approach. The theoretical basis of
such an approach rely on the assumption that at higher energies
beyond where the SM is valid, there is a more fundamental theory
which reduces to the SM at lower energies. The SM is assumed to be
an effective low-energy theory in which heavy fields have been
integrated out. Such a procedure is quite general and independent of
the model at the new physics energy scale.

Specifically we consider the $SU(2)_L\otimes U(1)_Y$ invariant
effective Lagrangian introduced in Refs.
\cite{FloresTlalpa:2001sp,DiazCruz:1999xe,Huang:2000st}. The
effective Lagrangian can be written as
\begin{eqnarray}
\label{effectivelag} {\cal L}_{eff}={\cal
L}_{SM}+\frac{1}{\Lambda^2}\sum_{n,i,j}c^{ij}_n O^{ij}_n +...&
(dim>6)
\end{eqnarray}
where $i,j=1,2,3$ denote flavor indices, $n$ runs over the number of
independent operators of dimension-6 and $\Lambda$ is the energy
scale of new physics. $c^{ij}_n$ are the dimensionless anomalous
coupling constants. There are four independent operators of
dimension-6 contributing to $W\ell_i \nu_j$ vertex:
\begin{eqnarray}
\label{eop1} O^{ij}_{LW}=(\bar L_i\gamma^\mu \tau^a D^\nu
L_j)W^a_{\mu\nu}+h.c.
\end{eqnarray}
\begin{eqnarray}
\label{eop2} O^{(3)ij}_{\phi L}=i(\phi^\dagger \tau^a D_\mu
\phi)(\bar L_i\gamma^\mu \tau^a L_j)+h.c.
\end{eqnarray}
\begin{eqnarray}
\label{eop3} O^{ij}_{D\ell}=(\bar L_i D_\mu \ell_{Rj})D^\mu\phi+h.c.
\end{eqnarray}
\begin{eqnarray}
\label{eop4} O^{ij}_{\ell W\phi}=\bar
L_i\sigma^{\mu\nu}\tau^a\ell_{Rj}\phi W^a_{\mu\nu}+h.c.
\end{eqnarray}
where $L_i$ and $\ell_{Rj}$ are the left-handed lepton doublet and
right-handed singlet of $SU(2)_L\otimes U(1)_Y$. $\phi$ is the
scalar doublet and $D_\mu$ is the covariant derivative.
$W^a_{\mu\nu}$ and $\tau^a$ are the $SU(2)_L$ gauge boson field
tensors and Pauli matrices respectively.

After spontaneous symmetry breaking, operators
(\ref{eop1}-\ref{eop4}) give rise to vertex functions for $W\ell_i
\nu_j$. The vertex functions for $W(k)\ell_i(p2) \nu_j(p1)$
generated from the effective Lagrangian are given, respectively, by
\begin{eqnarray}
\label{vertex1} \Gamma^\lambda_{(1)}=c_{1} \sqrt 2\left[-p_1^\lambda
k_\mu\gamma^\mu P_L +p_2^\lambda k_\mu\gamma^\mu P_L -(k\cdot
p_2)\gamma^\lambda P_L +(k\cdot p_1)\gamma^\lambda P_L\right]
\end{eqnarray}
\begin{eqnarray}
\label{vertex2} \Gamma^\lambda_{(2)}=-c_{2}\frac{g\eta^2}{\sqrt
2}\gamma^\lambda P_L
\end{eqnarray}
\begin{eqnarray}
\label{vertex3}
\Gamma^\lambda_{(3)}=-c_{3}\frac{g\eta}{2}p_2^\lambda P_L
\end{eqnarray}
\begin{eqnarray}
\label{vertex4} \Gamma^\lambda_{(4)}=2i\eta c_{4} k_\rho
\sigma^{\rho\lambda} P_L
\end{eqnarray}
For a convention, we assume that $p_1$ is incoming to and $p_2$ and
$k$ are outgoing from the vertex. $c_i$ $(i=1,..,4)$ are scaled
coupling constants defined by: $c_1=\frac{c_{LW}}{\Lambda^2}$,
$c_2=\frac{c^{(3)}_{\phi L}}{\Lambda^2}$,
$c_3=\frac{c_{D\ell}}{\Lambda^2}$ and $c_4=\frac{c_{\ell W
\phi}}{\Lambda^2}$ (For abbreviation we drop flavor indices).
$P_L=\frac{1}{2}(1-\gamma_5)$, $g$ is $SU(2)_L$ coupling constant
and $\eta$ represents the vacuum expectation value of the scalar
field. (For definiteness, we take $\eta=246$ GeV in the calculations
presented in this paper).

Operators (\ref{eop1}-\ref{eop4}) not only contribute to $W\ell_i
\nu_j$ but also $Z\ell_i^- \ell_j^+$ couplings. On the other hand,
$Z\ell_i^- \ell_j^+$ receive contributions from 9 independent
operators \cite{Huang:2000st}. Therefore processes involving
$Z\ell_i^- \ell_j^+$ vertex do not isolate the contributions coming
from operators (\ref{eop1}-\ref{eop4}).

\section{Neutrino fluxes and cross sections}

In a beta beam facility very intense and collimated neutrino or
antineutrino beams can be produced by accelerating $\beta$-unstable
heavy ions to a given $\gamma$ factor and allowing them to decay in
the straight section of a storage ring. In the ion rest frame the
neutrino spectrum is given by the following formula
\begin{eqnarray}
\frac{dN}{d\cos\theta dE_\nu}\sim
E_\nu^2(E_0-E_\nu)\sqrt{(E_\nu-E_0)^2-m_e^2}
\end{eqnarray}
where $E_0$ is the electron end-point energy, $m_e$ is the electron
mass. $E_\nu$ and $\theta$ are the energy and polar angle of the
neutrino. Neutrino flux observed in the laboratory frame can be
obtained by performing a Lorentz boost. The neutrino flux per solid
angle in a detector located at a distance $L$ is given by
\cite{BurguetCastell:2003vv}
\begin{eqnarray}
\left(\frac{d\phi^{Lab}}{dSdy}\right)_{\theta \simeq0}\simeq
\frac{N_\beta}{\pi
L^2}\frac{\gamma^2}{g(y_e)}y^2(1-y)\sqrt{(1-y)^2-y_e^2} ,
\end{eqnarray}
where $0\leq y\leq1-y_e$, $y=\frac{E_\nu}{2\gamma E_0}$,
$y_e=\frac{m_e}{E_0}$  and
\begin{eqnarray}
g(y_e)=\frac{1}{60}\left(\sqrt{1-y_e^2}(2-9y_e^2-8y_e^4)+15y_e^4Log\left[\frac{y_e}{1-\sqrt{1-y_e^2}}\right]
\right) .
\end{eqnarray}
$^{18}Ne$ and $^6He$ ions have been proposed as ideal candidates for
a neutrino and an antineutrino source, respectively
\cite{Zucchelli:2002sa,BurguetCastell:2003vv}. These ions produce
pure (anti)neutrino beams via the reactions
$^{18}_{10}Ne\to^{18}_9Fe^+\nu_e$ and $^6_2He^{++}\to^6_3Li^{+++}e^-
\bar{\nu}_e$. We assume that total number of ion decays per year is
$N_\beta=1.1\times10^{18}$ for $^{18}Ne$ and
$N_\beta=2.9\times10^{18}$ for $^6He$.

\begin{figure}
\includegraphics{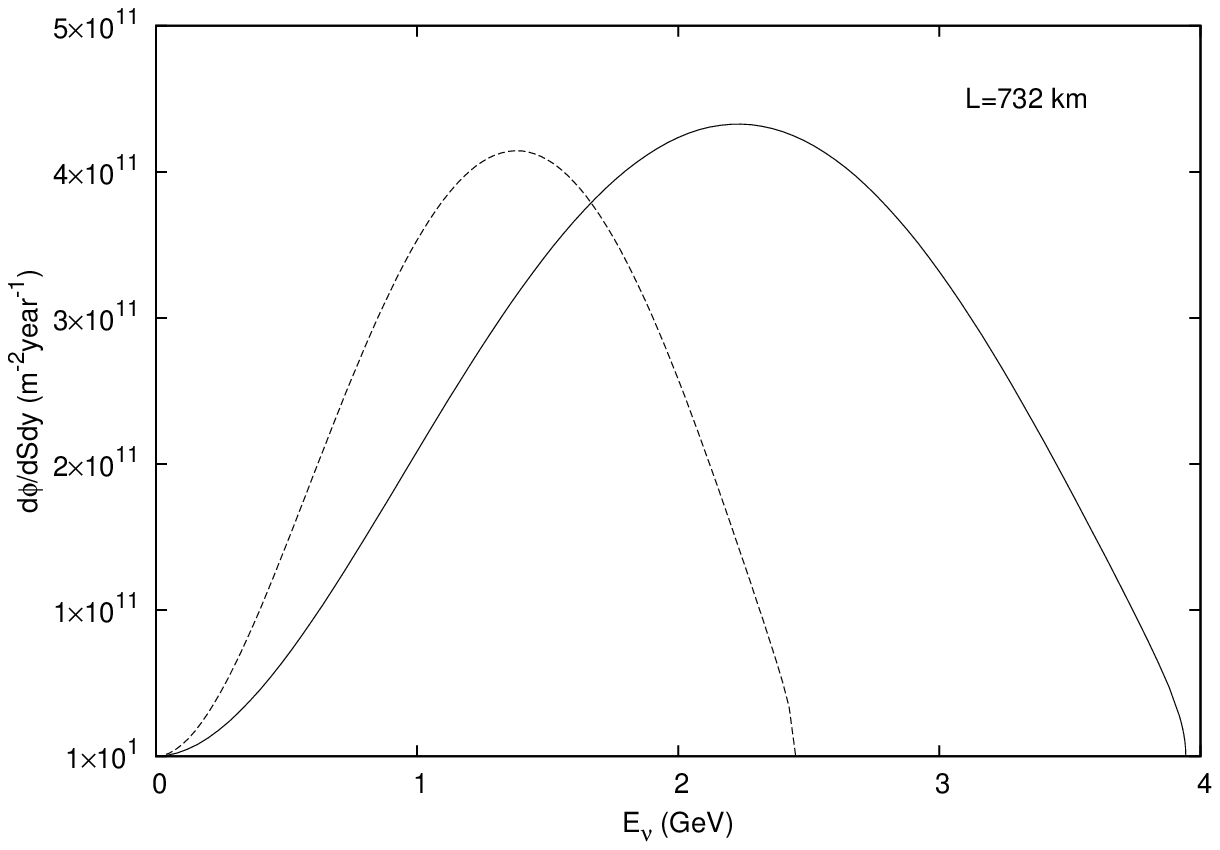}
\caption{Fluxes as a function of neutrino energy for $\nu_e$ (solid
line) and $\bar{\nu}_e$ (dotted line) at a detector of L = 732 km
distance. $\gamma$ parameter is taken to be 350 for $\bar{\nu}_e$
and 580 for $\nu_e$.\label{fig1}}
\end{figure}

Neutrino and antineutrino fluxes as a function of (anti)neutrino
energy at a detector of L = 732 km distance are plotted in Fig.
\ref{fig1}. $\gamma$ parameters for ions are taken to be
$\gamma=350$ for $^6He$ and $\gamma=580$ for $^{18}Ne$. The
foregoing detector distance and $\gamma$ values have been proposed
in Ref. \cite{BurguetCastell:2003vv} as a medium energy setup. In
Ref. \cite{BurguetCastell:2003vv} authors have considered a
Megaton-class water Cerenkov detector with a fiducial mass of 400
kiloton. They show that a cut demanding the reconstructed energy to
be larger than 500 MeV suppresses most of the residual backgrounds.
We assumed a water Cerenkov detector with the same mass and a cut of
500 MeV for the calculations presented in this paper.

In Fig.\ref{fig1} we observe that neutrino spectrum extend up to 4
GeV and antineutrino spectrum extend up to 2.5 GeV. The energy range
of the neutrino spectrum is comparably larger than the antineutrino
spectrum. Therefore number of events for antineutrinos is expected
to be low. So we do not perform an  analysis for antineutrinos. For
neutrino energies between 0.5 - 1.5 GeV, dominant contribution to
the cross section is provided by quasielastic scattering. When
neutrino energy exceeds 1.5 GeV, deep inelastic scattering starts to
dominate the cross section. Neutrino electron scattering takes place
at all energies in the spectrum. Lepton flavor violating $\nu_e e^-
\to \nu_e \mu^-$ process contains both $W$ and $Z$ exchange
diagrams. Hence this process receives contributions from both $W\mu
\nu_e$ and $Z\mu e$ couplings. Therefore it is not possible to set
limits on $W\mu \nu_e$ coupling independent from $Z\mu e$. Since the
main advantage of a beta beam facility has been lost (isolation of
the $W\mu \nu_e$ vertex) we do not analyse the process $\nu_e e^-
\to \nu_e \mu^-$.

\subsection{Neutrino nucleon quasielastic scattering}

\begin{figure}
\includegraphics{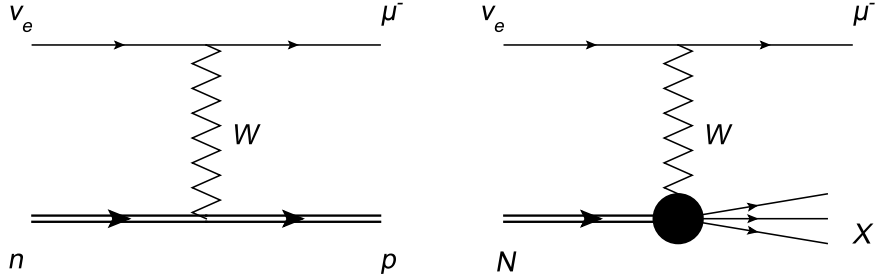}
\caption{Figure on the left shows Feynman diagram for $\nu_e n\to
\mu^-\, p$. Figure on the right represents schematic diagram for
neutrino deep inelastic scattering $\nu_e N\to \mu^-
X$.\label{fig2}}
\end{figure}

In addition to SM quasielastic scattering $\nu_e n\to e^-\, p$,
operators (\ref{eop1}-\ref{eop4}) give rise to new reactions $\nu_e
n\to \mu^-\, p$ and $\nu_e n\to \tau^-\, p$. The tau production
threshold is 3.5 GeV. Hence, tau production calls for a higher
$\gamma$ factor and medium energy setup of a beta-beam facility is
not convenient to analyse $\nu_e n\to \tau^- p$ reaction. On the
other hand, muon production via quasielastic scattering seems to be
appealing. Feynman diagram for $\nu_e n\to \mu^-\, p$ is given in
the left panel of Fig.\ref{fig2}. We see from the diagram that
hadron current is given by the standard formula \cite{Fukugita},
\begin{eqnarray}
\label{hadroncurrent} J^\mu_h=\frac{g}{2\sqrt 2}\cos
\theta_C\,\bar{u}_p\left[\gamma^\mu F_V -\gamma^\mu \gamma_5
F_A+\frac{1}{2m_N}i\sigma^{\mu\nu}q_\nu F_W\right]u_n
\end{eqnarray}
where $\cos \theta_C=0.974$ is the Cabibbo angle, $m_N$ is the mass
of the nucleon and $F$'s are invariant form factors that depend on
the transferred momentum $q^2\equiv(p_p-p_n)^2$. The $F$'s are known
as vector $F_V$, axial-vector $F_A$ and tensor $F_W$ (or weak
magnetism) form factors. They are all G-parity invariant. We adopt
the same parameterization of the momentum dependence as in Ref.
\cite{Strumia:2003zx}:
\begin{eqnarray}
F_V(q^2)=\frac{1-\frac{(1+\xi)q^2}{4m_N^2}}{\left(1-\frac{q^2}{4m_N^2}\right)
\left(1-\frac{q^2}{(0.84\,GeV)^2}\right)^{2}} \nonumber\\
F_W(q^2)=\frac{\xi}{\left(1-\frac{q^2}{4m_N^2}\right)
\left(1-\frac{q^2}{(0.84\,GeV)^2}\right)^{2}}\\
F_A(q^2)=1.270\left(1-\frac{q^2}{(1.032\,GeV)^2}\right)^{-2}\nonumber
\end{eqnarray}
Here $\xi=(\mu_p-\mu_n)/\mu_N=3.706$ is the difference in the
anomalous magnetic moments of the nucleons. Lepton current generated
from the effective vertices (\ref{vertex1}-\ref{vertex4}) is given
by
\begin{eqnarray}
\label{leptoncurrent}
J^\lambda_\ell=\bar{u}_\mu\left[\Gamma^\lambda_{(1)}+\Gamma^\lambda_{(2)}
+\Gamma^\lambda_{(3)}+\Gamma^\lambda_{(4)}\right]u_{\nu_e}
\end{eqnarray}
From (\ref{hadroncurrent}) and (\ref{leptoncurrent}) scattering
amplitude is obtained as
\begin{eqnarray}
\label{amplitude} M=\frac{\left(g_{\mu\lambda}-\frac{q_\mu
q_\lambda}{m_W^2}\right)}{q^2-m_W^2}\, J^\mu_h J^\lambda_\ell
\end{eqnarray}
where $m_W$ is the W boson mass. The W propagator can be
approximated as; $\frac{g_{\mu\lambda}-\frac{q_\mu
q_\lambda}{m_W^2}}{q^2-m_W^2}\approx-\frac{g_{\mu\lambda}}{m_W^2}$

\begin{figure}
\includegraphics{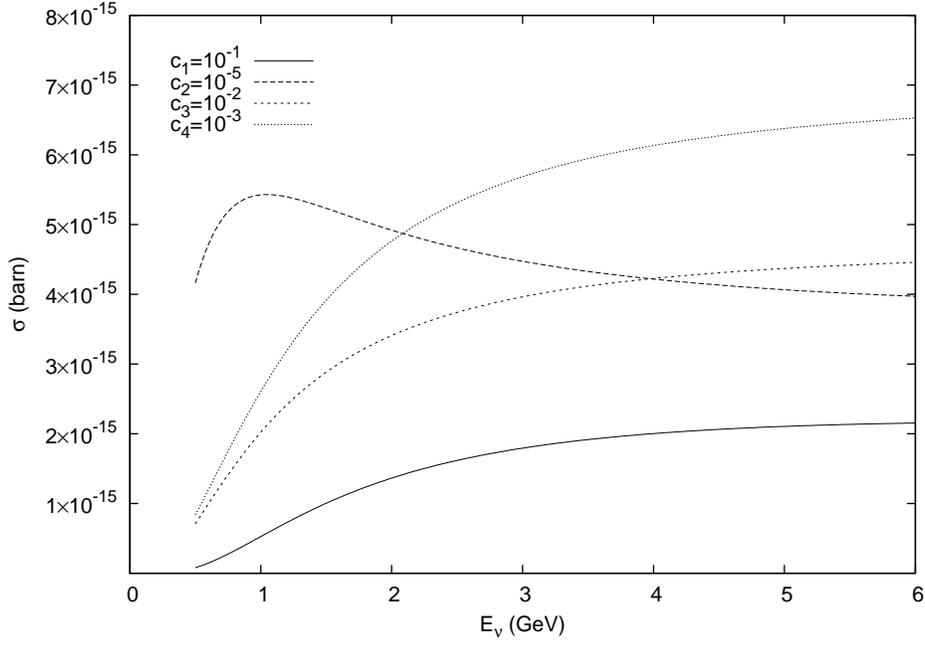}
\caption{Total cross section of $\nu_e n\to \mu^-\, p$ as a function
of neutrino energy. The values of the anomalous couplings is stated
in the figure. \label{fig3}}
\end{figure}

In Fig.\ref{fig3} we show total cross section of $\nu_e n\to \mu^-\,
p$ as a function of neutrino energy for some values of the anomalous
couplings. We see from the figure that cross sections proportional
to couplings $c_1,\,c_3$ and $c_4$ have similar behaviors as a
function of neutrino energy. They all increase as the energy
increases but increment rate is especially high up to 2 GeV. On the
other hand, anomalous cross section for $c_2$ attains its maximum at
approximately 1 GeV and then starts to decrease.

\subsection{Charged-current deep inelastic scattering}

When neutrino energy exceeds 1.5 GeV, deep inelastic scattering
starts to dominate the cross section. Charged-current deep inelastic
scattering of an electron-neutrino from nucleon is described by
t-channel $W$ exchange diagram (right panel of Fig.\ref{fig2}).
Since quark couplings to $W$ boson are not modified by operators
(\ref{eop1}-\ref{eop4}) hadron tensor does not receive any
contribution. It is defined in the standard form \cite{Particle Data
Group}
\begin{eqnarray}
W_{\mu\nu}=\left(-g_{\mu\nu}+\frac{q_\mu
q_\nu}{q^2}\right)F_1(x,Q^2)+\frac{\hat{p}_\mu \hat{p}_\nu}{p\cdot
q}F_2(x,Q^2)-i\epsilon_{\mu\nu\alpha\beta} \frac{q^\alpha
p^\beta}{2p\cdot q} F_3(x,Q^2)
\end{eqnarray}
where $p_\mu$ is the nucleon momentum, $q_\mu$ is the momentum of
the W boson propagator, $Q^2=-q^2$, $x=\frac{Q^2}{2p\cdot q}$ and
\begin{eqnarray}
\hat{p}_\mu\equiv p_\mu-\frac{p\cdot q}{q^2}q_\mu . \nonumber
\end{eqnarray}
The structure functions for scattering on a proton are defined as
follows \cite{Particle Data Group}
\begin{eqnarray}
\label{CC}
F_2^{W^+}=&&2x(d+\bar u+\bar c+s)\nonumber\\
F_3^{W^+}=&&2(d-\bar u-\bar c+s)
\end{eqnarray}
The form factors $F_1$'s can be obtained from (\ref{CC}) by using
Callan-Gross relation $2xF_1=F_2$ \cite{Callan:1969uq}. The
structure functions for scattering on a neutron are obtained from
those of the proton by the interchange $u\leftrightarrow d$. In our
calculations parton distribution functions of Martin {\it et
al.}\cite{pdf} have been used. We assumed an isoscalar oxygen
nucleus $N=(p+n)/2$ and two free protons for each $H_2O$ molecule.
Naturally occurring oxygen is 99.8\% $^{16}$O which is isoscalar
\cite{ti}. Hence the error incurred by assuming an isoscalar oxygen
target would be not more than a fraction of one percent.

Possible new physics contributions coming from the operators
(\ref{eop1}-\ref{eop4}) only modify the lepton tensor:
\begin{eqnarray}
\label{LeptonTensor} L_{\mu\nu}=\sum_{spin}\left[\bar u(k^\prime)
\Gamma_\mu u(k)\right]^\dagger \left[\bar u(k^\prime) \Gamma_\nu
u(k)\right]
\end{eqnarray}
where $k$ and $k^\prime$ are the momenta of initial $\nu_e$ and
final $\mu^-$, respectively. $\Gamma_\mu$ represents the sum of
anomalous vertices (\ref{vertex1}-\ref{vertex4}).

\begin{figure}
\includegraphics{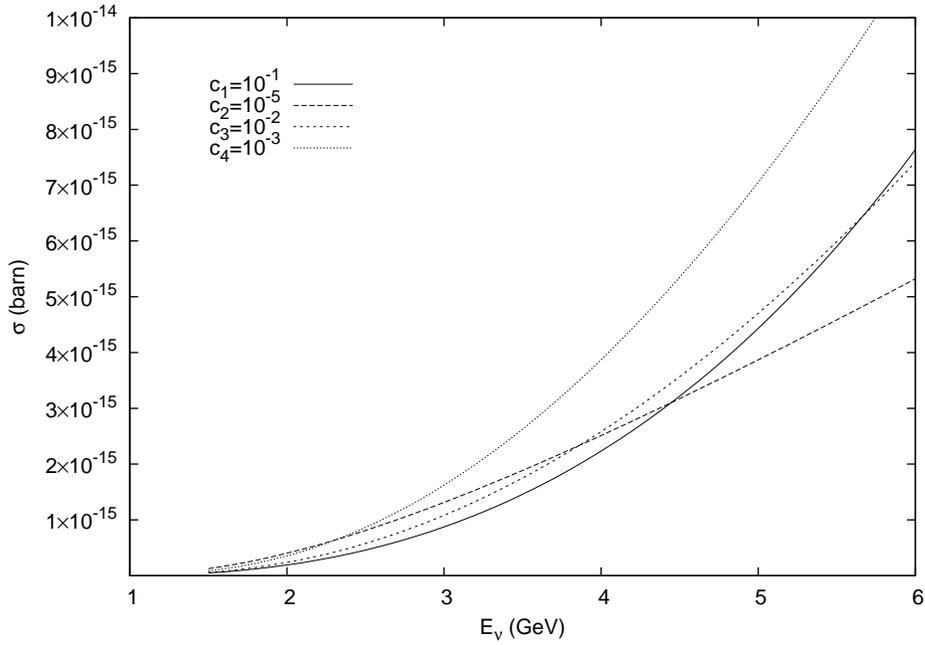}
\caption{Total cross section of charged-current deep inelastic
scattering $\nu_e p\to \mu^- X$ as a function of neutrino energy.
The values of the anomalous couplings is stated in the figure.
\label{fig4}}
\end{figure}

The muon production via charged-current deep inelastic scattering of
neutrinos from the proton is plotted in Fig.\ref{fig4}. We see from
the figure that different from quasielastic scattering case,
increment rate is higher at high energies.

\subsection{Statistical analysis and results}

\begin{figure}
\includegraphics{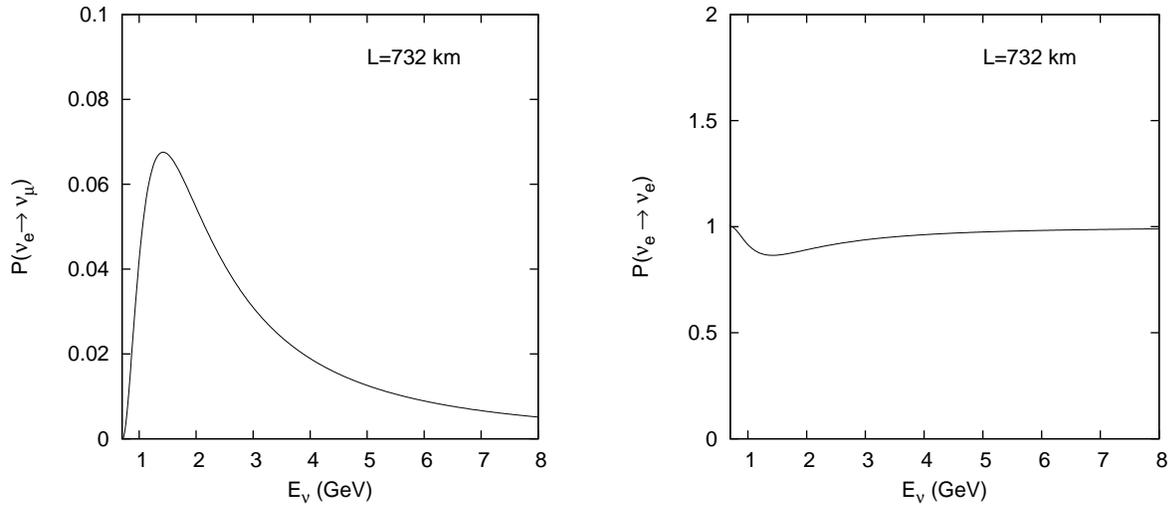}
\caption{Figure on the left is the transition probability $P(\nu_e
\to \nu_\mu)$ and figure on the right is the survival probability
$P(\nu_e \to \nu_e)$ in vacum as a function of neutrino energy.
Detector distance is taken to be 732 km. \label{fig5}}
\end{figure}

A detailed investigation of the anomalous couplings requires a
statistical analysis. To this purpose, number of events have to be
calculated. If we assume that initial neutrino beam is pure and
contains only electron-neutrinos then SM cross section for muon
production is zero. But neutrino beam near the detector should
contain a small fraction of muon-neutrinos due to neutrino
oscillations. In Fig.\ref{fig5} we present the transition
probability $P(\nu_e \to \nu_\mu)$ and the survival probability
$P(\nu_e \to \nu_e)$ as a function of neutrino energy at a detector
of distance 732 km. We use the following approximate formulas
\cite{Bilenky}
\begin{eqnarray}
\label{transition} P(\nu_e \to
\nu_\mu)=&&\frac{1}{2}\sin^2(2\theta_{13})\sin^2\theta_{23}\left(1-\cos\left(2.54\frac{\Delta
m^2_{23}L} {E_\nu}\right)\right)
\end{eqnarray}
\begin{eqnarray}
\label{survival} P(\nu_e \to
\nu_e)=&&1-\frac{1}{2}\sin^2(2\theta_{13})\left(1-\cos\left(2.54\frac{\Delta
m^2_{23}L} {E_\nu}\right)\right)
\end{eqnarray}
where $\Delta m^2_{23}$ is the neutrino mass-squared difference in
$eV^2$, L is the distance between neutrino source and detector in
$m$ and neutrino energy $E_\nu$ is given in units of $MeV$. In the
calculations we assume that $\sin^2\theta_{13}=0.035$,
$\sin^2\theta_{23}=0.50$ and $|\Delta
m^2_{23}|=2.40\times10^{-3}\,eV^2$ \cite{Schwetz:2008er}. Number of
events has been obtained by integrating cross section over the
neutrino energy spectrum and transition probability and multiplying
by the appropriate factor that accounts for the number of
corresponding particles (protons or neutrons) in a 400 kiloton
fiducial mass of the detector. For instance, number of SM events for
charged-current deep inelastic scattering is given through the
formula,
\begin{eqnarray}
N_{SM}=\int P(\nu_e \to \nu_\mu)
\left(\frac{d\phi^{Lab}}{dSdE_\nu}\right)[N_{p}\, \sigma_{\nu_\mu
p\to \mu^- X}+N_{n}\, \sigma_{\nu_\mu n\to \mu^- X}] dE_\nu
\end{eqnarray}
where $\sigma_{\nu_\mu p\to \mu^- X}$ and $\sigma_{\nu_\mu n\to
\mu^- X}$ are the SM cross sections of deep inelastic scattering of
the muon-neutrino from the proton and neutron respectively. $N_{p}$
and $N_{n}$ are the number of protons and neutrons in a 400 kiloton
fiducial mass of the detector. Number of lepton flavor violating
events can be calculated in a similar manner but considering lepton
flavor violating cross sections and survival probability.

We studied $95\%$ confidence level (C.L.) bounds using one-parameter
$\chi^{2}$ analysis without a systematic error. The $\chi^{2}$
function is given by,
\begin{eqnarray}
\chi^{2}=\left(\frac{N}{N_{SM} \,\, \delta_{stat}}\right)^{2}
\end{eqnarray}
where $N$ is the number of lepton flavor violating events as a
function of couplings $c_1,c_2,c_3,c_4$, $N_{SM}$ is the number of
events expected in the SM and $\delta_{stat}$ is the statistical
error. In Table \ref{tab1}, we show $95\%$ C.L. bounds on the
couplings $c_1,c_2,c_3$ and $c_4$ obtained from quasielastic and
deep inelastic scatterings. These bounds are obtained for 1 year
running time of the beta-beam experiment. We see from the table that
bounds obtained from quasielastic and deep inelastic scatterings are
close to each other. Therefore, they have almost same potential to
probe LFV.

\section{Conclusions}
Beta beams present an ideal venue to measure neutrino cross
sections. For beta beams neutrino fluxes are precisely known and
therefore uncertainties associated with the neutrino (antineutrino)
fluxes are negligible. Purity of the produced (anti)
electron-neutrino beam is an other advantage of beta beams. These
features enable to detect neutrino cross sections with a high
precision.

Medium energy setup of a beta-beam facility provides us the
opportunity to isolate $W\mu\nu_e$ vertex which is not the case for
a collider experiment. Probing $W\mu\nu_e$ couplings is important
for understanding the physics beyond the SM and contributes to the
studies in neutrino physics. In neutrino oscillation experiments,
identification of the neutrino flavor is based on charged-current
processes at the detector. Neutrino flavor can be identified through
the flavor of the associated charged lepton. Hence, lepton flavor
violating couplings of the neutrinos to $W$ can mimic the
oscillation signals. Therefore constraints on $W\mu\nu_e$ couplings
are important for precision measurements of the oscillation
parameters.


\pagebreak

\begin{table}
\caption{$95\%$ C.L. bounds on the couplings $c_1,c_2,c_3$ and $c_4$
obtained from quasielastic and deep inelastic scatterings. Only one
of the couplings is assumed to be non-zero at a time.\label{tab1}}
\begin{ruledtabular}
\begin{tabular}{ccccc}
Quasielastic scattering:& & & &  \\
&$|c_1|<0.0159$ &$|c_2|<8.457\times10^{-7}$ &$|c_3|<0.0010$ &$|c_4|<8.561\times 10^{-5}$  \\
\hline
Deep inelastic scattering:& & & & \\
&$|c_1|<0.0105$ &$|c_2|<8.902\times10^{-7}$ &$|c_3|<0.0010$ &$|c_4|<7.846\times 10^{-5}$  \\
\end{tabular}
\end{ruledtabular}
\end{table}


\begin{thebibliography}{99}

\bibitem{Kamiokande1} Y. Fukuda {\it et al.}, Phys. Lett. B {\bf 335}, 237
(1994).

\bibitem{Kamiokande2} Y. Fukuda {\it et al.}, Phys. Rev. Lett. {\bf 81}, 1562 (1998).

\bibitem{Ahmad:2002jz}
  Q.~R.~Ahmad {\it et al.},
  Phys.\ Rev.\ Lett.\  {\bf 89}, 011301 (2002)
  [nucl-ex/0204008].

\bibitem{DiazCruz:1999xe}
  J.~L.~Diaz-Cruz and J.~J.~Toscano,
  Phys.\ Rev.\  D {\bf 62}, 116005 (2000)
  [arXiv:hep-ph/9910233].

\bibitem{FloresTlalpa:2001sp}
  A.~Flores-Tlalpa, J.~M.~Hernandez, G.~Tavares-Velasco and J.~J.~Toscano,
  Phys.\ Rev.\  D {\bf 65}, 073010 (2002)
  [arXiv:hep-ph/0112065].

\bibitem{Zucchelli:2002sa}
  P.~Zucchelli,
  Phys.\ Lett.\  B {\bf 532} (2002) 166.

\bibitem{Volpe:2003fi}
  C.~Volpe,
  J.\ Phys.\ G {\bf 30} (2004) L1
  [arXiv:hep-ph/0303222].

\bibitem{McLaughlin:2003yg}
  G.~C.~McLaughlin and C.~Volpe,
  Phys.\ Lett.\  B {\bf 591} (2004) 229
  [arXiv:hep-ph/0312156].

\bibitem{Serreau:2004kx}
  J.~Serreau and C.~Volpe,
  Phys.\ Rev.\  C {\bf 70} (2004) 055502
  [arXiv:hep-ph/0403293].

\bibitem{McLaughlin:2004va}
  G.~C.~McLaughlin,
  Phys.\ Rev.\  C {\bf 70} (2004) 045804
  [arXiv:nucl-th/0404002].

\bibitem{Volpe:2005iy}
  C.~Volpe,
  J.\ Phys.\ G {\bf 31} (2005) 903
  [arXiv:hep-ph/0501233].

\bibitem{Balantekin:2005md}
  A.~B.~Balantekin, J.~H.~de Jesus and C.~Volpe,
  Phys.\ Lett.\  B {\bf 634} (2006) 180
  [arXiv:hep-ph/0512310].

\bibitem{Balantekin:2006ga}
  A.~B.~Balantekin, J.~H.~de Jesus, R.~Lazauskas and C.~Volpe,
  Phys.\ Rev.\  D {\bf 73} (2006) 073011
  [arXiv:hep-ph/0603078].

\bibitem{Volpe:2006in}
  C.~Volpe,
  J.\ Phys.\ G {\bf 34} (2007) R1
  [arXiv:hep-ph/0605033].

\bibitem{Jachowicz:2006xx}
  N.~Jachowicz and G.~C.~McLaughlin,
  Phys.\ Rev.\ Lett.\  {\bf 96} (2006) 172301
  [arXiv:nucl-th/0604046].

\bibitem{Bueno:2006yq}
  A.~Bueno, M.~C.~Carmona, J.~Lozano and S.~Navas,
  Phys.\ Rev.\  D {\bf 74} (2006) 033010.

\bibitem{Lazauskas:2007va}
  R.~Lazauskas, A.~B.~Balantekin, J.~H.~De Jesus and C.~Volpe,
  Phys.\ Rev.\  D {\bf 76} (2007) 053006
  [arXiv:hep-ph/0703063].

\bibitem{Amanik:2007zy}
  P.~S.~Amanik and G.~C.~McLaughlin,
  Phys.\ Rev.\  C {\bf 75} (2007) 065502
  [arXiv:hep-ph/0702207].

\bibitem{Jachowicz:2008kx}
  N.~Jachowicz, G.~C.~McLaughlin and C.~Volpe,
  Phys.\ Rev.\  C {\bf 77} (2008) 055501
  [arXiv:0804.0360 [nucl-th]].

\bibitem{BurguetCastell:2003vv}
  J.~Burguet-Castell, D.~Casper, J.~J.~Gomez-Cadenas, P.~Hernandez and F.~Sanchez,
  Nucl.\ Phys.\  B {\bf 695} (2004) 217
  [arXiv:hep-ph/0312068].

\bibitem{BurguetCastell:2005pa}
  J.~Burguet-Castell, D.~Casper, E.~Couce, J.~J.~Gomez-Cadenas and P.~Hernandez,
  Nucl.\ Phys.\  B {\bf 725} (2005) 306
  [arXiv:hep-ph/0503021].

\bibitem{Migliozzi:2005zh}
  P.~Migliozzi,
  Nucl.\ Phys.\ Proc.\ Suppl.\  {\bf 145} (2005) 199.

\bibitem{Donini:2006tt}
  A.~Donini, E.~Fernandez-Martinez, P.~Migliozzi, S.~Rigolin, L.~Scotto Lavina, T.~Tabarelli de Fatis and F.~Terranova,
  Eur.\ Phys.\ J.\  C {\bf 48} (2006) 787
  [arXiv:hep-ph/0604229].

\bibitem{Donini:2005qg}
  A.~Donini, E.~Fernandez, P.~Migliozzi, S.~Rigolin, L.~Scotto Lavina, T.~Tabarelli de Fatis and F.~Terranova,
  arXiv:hep-ph/0511134.

\bibitem{Donini:2004hu}
  A.~Donini, E.~Fernandez-Martinez, P.~Migliozzi, S.~Rigolin and L.~Scotto Lavina,
  Nucl.\ Phys.\  B {\bf 710}  (2005) 402
  [arXiv:hep-ph/0406132].

\bibitem{Terranova:2004hu}
  F.~Terranova, A.~Marotta, P.~Migliozzi and M.~Spinetti,
  Eur.\ Phys.\ J.\  C {\bf 38} (2004) 69
  [arXiv:hep-ph/0405081].

\bibitem{Huber:2005jk}
  P.~Huber, M.~Lindner, M.~Rolinec and W.~Winter,
  Phys.\ Rev.\  D {\bf 73} (2006) 053002
  [arXiv:hep-ph/0506237].

\bibitem{Mezzetto:2003ub}
  M.~Mezzetto,
  J.\ Phys.\ G {\bf 29} (2003) 1771
  [arXiv:hep-ex/0302007].

\bibitem{Balantekin:2008vw}
  A.~B.~Balantekin, I.~Sahin, B.~Sahin,
  JHEP {\bf 0905}, 005 (2009)
  [arXiv:0812.1722 [hep-ph]].

\bibitem{Agarwalla:2006ht}
  S.~K.~Agarwalla, S.~Rakshit, A.~Raychaudhuri,
  Phys.\ Lett.\  {\bf B647}, 380-388 (2007)
  [hep-ph/0609252].


\bibitem{Davidson:2003ha}
  S.~Davidson, C.~Pena-Garay, N.~Rius and A.~Santamaria,
  JHEP {\bf 0303} (2003) 011
  [arXiv:hep-ph/0302093].

\bibitem{Bell:2005kz}
  N.~F.~Bell, V.~Cirigliano, M.~J.~Ramsey-Musolf, P.~Vogel and M.~B.~Wise,
  Phys.\ Rev.\ Lett.\  {\bf 95} (2005) 151802
  [arXiv:hep-ph/0504134].

\bibitem{Ibarra:2004pe}
  A.~Ibarra, E.~Masso, J.~Redondo,
  Nucl.\ Phys.\  {\bf B715}, 523-535 (2005).
  [hep-ph/0410386].

\bibitem{Mohapatra:2005wg}
  R.~N.~Mohapatra {\it et al.},
  Rept.\ Prog.\ Phys.\  {\bf 70} (2007) 1757
  [arXiv:hep-ph/0510213].

\bibitem{Mohapatra:2006gs}
  R.~N.~Mohapatra and A.~Y.~Smirnov,
  Ann.\ Rev.\ Nucl.\ Part.\ Sci.\  {\bf 56} (2006) 569
  [arXiv:hep-ph/0603118].

\bibitem{de Gouvea:2007xp}
  A.~de Gouvea and J.~Jenkins,
  Phys.\ Rev.\  D {\bf 77} (2008) 013008
  [arXiv:0708.1344 [hep-ph]].

\bibitem{Perez:2008ha}
  P.~Fileviez Perez, T.~Han, G.~Y.~Huang, T.~Li and K.~Wang,
  Phys.\ Rev.\  D {\bf 78} (2008) 015018
  [arXiv:0805.3536 [hep-ph]].

\bibitem{Balantekin:2008rc}
  A.~B.~Balantekin, I.~Sahin and B.~Sahin,
  Phys.\ Rev.\  D {\bf 78} (2008) 073003
  [arXiv:0807.3385 [hep-ph]].

\bibitem{Balantekin:2008ib}
  A.~B.~Balantekin and I.~Sahin,
   J.\ Phys.\ G {\bf 36} (2009) 025010
  [arXiv:0810.4318 [hep-ph]].

\bibitem{Biggio:2009nt}
  C.~Biggio, M.~Blennow and E.~Fernandez-Martinez,
  JHEP {\bf 0908}, 090 (2009)
  [arXiv:0907.0097 [hep-ph]].


\bibitem{Huang:2000st}
  T.~Huang, Z.~H.~Lin, X.~Zhang,
  [hep-ph/0009353].

\bibitem{Fukugita}
   M. Fukugita and T. Yanagida, {\em Physics of Neutrinos and Applications to
    Astrophysics} (Springer-Verlag, Berlin, 2003).

\bibitem{Strumia:2003zx}
  A.~Strumia, F.~Vissani,
  Phys.\ Lett.\  {\bf B564}, 42-54 (2003)
  [astro-ph/0302055].

\bibitem{Particle Data Group} Particle Data Group Collaboration, K.
Nakamura {\it et al}, J. Phys. G {\bf 37}, 075021 (2010).

\bibitem{Callan:1969uq}
  C.~G.~Callan and D.~J.~Gross,
  Phys.\ Rev.\ Lett.\  {\bf 22} (1969) 156.

\bibitem{pdf}  A. D. Martin, W. J. Stirling, R. S. Thorne and G.
Watt, Phys. Lett. B {\bf 652}, 292 (2007) [arXiv:0706.0459
[hep-ph]].

\bibitem{ti}
R.B. Firestone, V.S. Shirley, {\em Table of Isotopes}, R.B.
Firestone, V.S. Shirley Eds. (Wiley, New York, 1996).

\bibitem{Bilenky} S. Bilenky, {\em Introduction to the Physics of Massive and Mixed Neutrinos}
(Springer-Verlag, Berlin, 2010).

\bibitem{Schwetz:2008er}
  T.~Schwetz, M.~A.~Tortola and J.~W.~F.~Valle,
  New J.\ Phys.\  {\bf 10}, 113011 (2008)
  [arXiv:0808.2016 [hep-ph]].

\end{thebibliography}
\end{document}